\documentclass[aps,prc,twocolumn,showpacs,superscriptaddress,amsmath,amssymb,nofootinbib]{revtex4}

\usepackage{graphicx}
\usepackage{dcolumn}
\usepackage{bm}
\begin{document}


\title{Spectroscopy of $^{24}$Al and extraction of Gamow-Teller strengths with the $^{24}$Mg($^{3}$He,$t$) reaction at 420 MeV.}

\author{R.G.T. Zegers}
\email{zegers@nscl.msu.edu} \affiliation{National Superconducting Cyclotron Laboratory, Michigan
State University, East Lansing, MI 48824-1321, USA} \affiliation{Department of Physics and
Astronomy, Michigan State University, East Lansing, MI 48824, USA} \affiliation{Joint Institute for
Nuclear Astrophysics, Michigan State University, East Lansing, MI 48824, USA}
\author{R. Meharchand}
\affiliation{National Superconducting Cyclotron Laboratory, Michigan State University, East
Lansing, MI 48824-1321, USA} \affiliation{Department of Physics and Astronomy, Michigan State
University, East Lansing, MI 48824, USA} \affiliation{Joint Institute for Nuclear Astrophysics,
Michigan State University, East Lansing, MI 48824, USA}
\author{T. Adachi}
\affiliation{Research Center for Nuclear Physics, Osaka University, Ibaraki, Osaka 567-0047, Japan}
\author{Sam M. Austin}
\affiliation{National Superconducting Cyclotron Laboratory, Michigan State University, East Lansing, MI 48824-1321, USA}
\affiliation{Joint Institute for Nuclear Astrophysics, Michigan State University, East Lansing, MI 48824, USA}
\author{B.A. Brown}
\affiliation{National Superconducting Cyclotron Laboratory, Michigan State University, East
Lansing, MI 48824-1321, USA} \affiliation{Department of Physics and Astronomy, Michigan State
University, East Lansing, MI 48824, USA} \affiliation{Joint Institute for Nuclear Astrophysics,
Michigan State University, East Lansing, MI 48824, USA}
\author{Y. Fujita}
\affiliation{Department of Physics, Osaka University, Toyonaka, Osaka 560-0043, Japan}
\author{M. Fujiwara}
\affiliation{Research Center for Nuclear Physics, Osaka University, Ibaraki, Osaka 567-0047, Japan}
\affiliation{Kansai Photon Science Institute, Japan Atomic Energy Agency, Kizu, Kyoto 619-0215,
Japan}
\author{C. J. Guess}
\affiliation{National Superconducting Cyclotron Laboratory, Michigan State University, East
Lansing, MI 48824-1321, USA} \affiliation{Department of Physics and Astronomy, Michigan State
University, East Lansing, MI 48824, USA} \affiliation{Joint Institute for Nuclear Astrophysics,
Michigan State University, East Lansing, MI 48824, USA}
\author{H. Hashimoto}
\affiliation{Research Center for Nuclear Physics, Osaka University, Ibaraki, Osaka 567-0047, Japan}
\author{K. Hatanaka}
\affiliation{Research Center for Nuclear Physics, Osaka University, Ibaraki, Osaka 567-0047, Japan}
\author{M.E. Howard}
\affiliation{Joint Institute for Nuclear Astrophysics, Michigan State University, East Lansing, MI
48824, USA} \affiliation{Department of Physics, The Ohio State University, Columbus, OH 43210, USA}
\author{H. Matsubara}
\affiliation{Research Center for Nuclear Physics, Osaka University, Ibaraki, Osaka 567-0047, Japan}
\author{K. Nakanishi}
\affiliation{Research Center for Nuclear Physics, Osaka University, Ibaraki, Osaka 567-0047, Japan}
\author{T. Ohta}
\affiliation{Research Center for Nuclear Physics, Osaka University, Ibaraki, Osaka 567-0047, Japan}
\author{H. Okamura}
\affiliation{Research Center for Nuclear Physics, Osaka University, Ibaraki, Osaka
567-0047, Japan}
\author{Y. Sakemi}
\affiliation{Research Center for Nuclear Physics, Osaka University, Ibaraki, Osaka 567-0047, Japan}
\author{Y. Shimbara}
\altaffiliation{Current Address: Graduate School of Science and Technology, Niigata University, Niigata 950-2181, Japan}
\affiliation{National Superconducting Cyclotron Laboratory, Michigan State University, East
Lansing, MI 48824-1321, USA} \affiliation{Joint Institute for Nuclear Astrophysics, Michigan State
University, East Lansing, MI 48824, USA}
\author{Y. Shimizu}
\affiliation{Research Center for Nuclear Physics, Osaka University, Ibaraki, Osaka 567-0047, Japan}
\author{C. Scholl}
\affiliation{Institut f\"ur Kernphysik, Universit\"at zu K\"oln, D-50937 Cologne, Germany}
\author{A. Signoracci}
\affiliation{National Superconducting Cyclotron Laboratory, Michigan State University, East
Lansing, MI 48824-1321, USA} \affiliation{Department of Physics and Astronomy, Michigan State
University, East Lansing, MI 48824, USA}
\author{Y. Tameshige}
\affiliation{Research Center for Nuclear Physics, Osaka University, Ibaraki, Osaka 567-0047, Japan}
\author{A. Tamii}
\affiliation{Research Center for Nuclear Physics, Osaka University, Ibaraki, Osaka 567-0047, Japan}
\author{M. Yosoi}
\affiliation{Research Center for Nuclear Physics, Osaka University, Ibaraki, Osaka 567-0047, Japan}
\date{\today}%

\begin{abstract}
The $^{24}$Mg($^{3}$He,$t$)$^{24}$Al reaction has been studied at $E(^{3}$He$)=420$ MeV. An energy resolution of 35 keV was achieved. Gamow-Teller strengths to discrete levels in $^{24}$Al are extracted by using a recently developed empirical relationship for the proportionality between Gamow-Teller strengths and differential cross sections at zero momentum transfer. Except for small discrepancies for a few weak excitations, good agreement with previous $^{24}$Mg($p,n$) data and nuclear-structure calculations using the USDA/B interactions in the $sd$ shell-model space is found. The excitation energy of several levels in $^{24}$Al of significance for determination of the $^{23}$Mg($p,\gamma)^{24}$Al thermonuclear reaction rate were measured. Results are consistent with two of the three previous ($^{3}$He,$t$) measurements, performed at much lower beam energies. However, a new state at $E_{x}$($^{24}$Al)=2.605(10) MeV was found and is the third state above the proton separation energy.
\end{abstract}

\pacs{21.60.Cs, 25.40.Kv, 25.55.Kr, 26.30.-k, 27.30.+t}
\maketitle

\section{Introduction}
\label{sec:intro}
Charge-exchange (CE) reactions with hadronic probes are an excellent tool for studying the spin-isospin response in nuclei \cite{HAR01,OST92} and specifically for extracting Gamow-Teller (GT; $\Delta L=0$, $\Delta S=1$, $\Delta T=1$) strength distributions. Since experimental $\beta$ decay studies only provide access to the GT response at low excitation energies, CE studies have become the preferred technique for mapping the more complete GT strength distribution.

The ($^{3}$He,$t$) reaction at 420-450 MeV has been used extensively for extracting GT strengths in the $\Delta T_{z}=-1$ direction (see e.g. Refs. \cite{FUJ96,FUJ00,FUJ04a,FUJ04b,FUJ05,ZEG06,FUJ07a}). Resolutions of as low as 20 keV in full-width at half-maximum (FWHM) have been achieved \cite{FUJ07}. Such high level of detail provides excellent testing ground for theoretical nuclear structure calculations. In addition, it allows for a relatively clean extraction of GT strengths from transitions with different angular momentum transfer. In the recent work of Ref. \cite{ZEG07}, an empirical mass-dependent relationship for the proportionality constant between GT strength and differential cross section at zero momentum transfer (the so-called unit cross section) was established, as had been done earlier for $(p,n)$ reactions \cite{TAD87}. This relationship is important for the extraction of GT strengths when the unit cross section cannot be directly obtained by using experimental $\beta$ decay log$ft$ values in the same nucleus, as is the case for GT transitions from $^{24}$Mg to $^{24}$Al discussed here. The GT strength distribution in $^{24}$Al has been obtained in the past, by using the $^{24}$Mg($p,n$) reaction at 135 MeV \cite{AND91}. In that analysis, GT strengths were extracted by using the empirical relationship for the unit cross section for the $(p,n)$ probe. A comparison between the ($^{3}$He,$t$) and $(p,n)$ results provides a good measure of the systematic errors made when employing these methods using different probes.

\begin{figure*}
\includegraphics[scale=0.9]{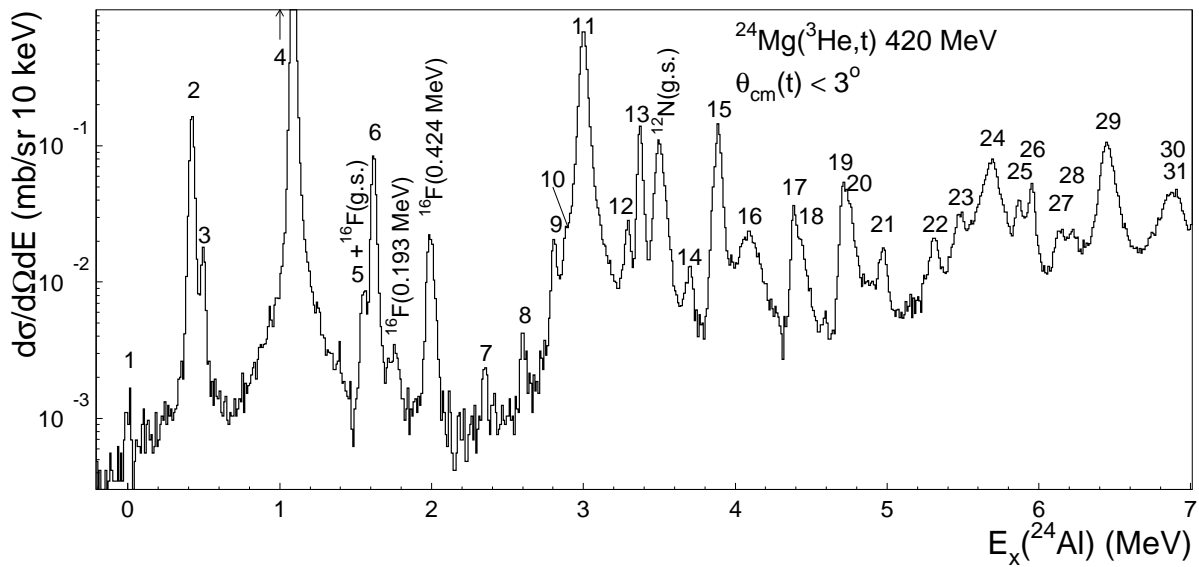}
\caption{\label{fig1} Energy spectrum of the $^{24}$Mg($^{3}$He,$t$) reaction, integrated over the full opening angle used in the analysis. Peaks identified as excited states in $^{24}$Al are numbered. Peaks due to contaminants in the target are labeled. Note that the vertical-axis scale is logarithmic and that the maximum of peak 4 is beyond the maximum vertical scale.}
\end{figure*}

The accurate knowledge of the location of excited states in $^{24}$Al is also important for calculating the thermonuclear $^{23}$Mg($p,\gamma)^{24}$Al reaction rate of \cite{WAL81,WIE86,KUB95,HER98,VIS07}. This reaction rate plays a significant role in explosive hydrogen-burning stars (e.g. novae) when the temperature is sufficiently high ($0.1-2\times10^{9}$ K) for proton capture on $^{23}$Mg to compete with $^23$Mg $\beta$ decay. The total proton-capture rate consists of a resonant contribution, due to unbound compound nuclear states, and a non-resonant direct-capture contribution. The resonant contributions depend exponentially on the resonance energies, which must thus be known with high accuracy. In the relevant excitation-energy region, just above the proton threshold in $^{24}$Al (1.87 MeV), the available data on the resonance energies stems from three ($^{3}$He,$t$) experiments performed at 81 MeV \cite{GRE91}, 60 MeV \cite{KUB95} and 30 MeV \cite{VIS07}. Results from the latter two are consistent, but differ by about $30-50$ keV from the results of Ref. \cite{GRE91}. In Ref. \cite{HER98}, resonance energies based on the then-adopted values \cite{END98} (averages of Refs. \cite{GRE91,KUB95}) were used to calculate the thermonuclear $^{23}$Mg($p,\gamma)^{24}$Al reaction rates. In Ref. \cite{VIS07}, newly measured resonance energies were used, leading to an increase in the proton-capture rate by 5-20\%, depending on the temperature. In the most recent compilation \cite{FIR07} (performed before the results from Ref. \cite{VIS07} became available) the adopted excitation energies were those of Ref. \cite{GRE91}.
In light of this history, an additional measurement of the relevant energies is desirable. The present ($^{3}$He,$t$) data are taken at much higher beam energies than the previous measurements and at very forward angles. The sensitivity for transitions with small angular momentum transfer are, therefore, enhanced. This increases the probability of finding new  resonances in the region just above the proton threshold.

\section{Experiment and data extraction.}
\label{sec:exp}
The $^{24}$Mg($^{3}$He,$t$) experiment was performed at the Research Center for Nuclear Physics (RCNP), Osaka University, by using a 420 MeV $^{3}$He$^{2+}$ beam of $\sim 15$ pnA produced in the ring cyclotron. Scattered tritons from a 0.7-mg/cm$^{2}$ thick, 99.92\% isotopically pure $^{24}$Mg target were momentum analyzed in the Grand Raiden spectrometer \cite{FUJ99}. During storage and use in other experiments, the target had somewhat oxidized and also contained traces of $^{12}$C. These contaminants proved useful for the energy calibration of the spectrum.
An energy resolution of 35 keV (FWHM) was achieved by using the lateral dispersion-matching technique \cite{FUJ02}. The spectrometer was set at an angle of $0^{\circ}$. Differential cross sections up to 3$^{\circ}$ in the center-of-mass system could be extracted. To optimize the angular resolution, the angular dispersion matching technique was applied \cite{FUJ02} and the experiment was run in the over-focus mode \cite{FUJ01,ZEG04}. A resolution of $~0.2^{\circ}$ in laboratory scattering angle was achieved.

The $^{3}$He$^{2+}$ beam was stopped in a Faraday cup, placed at the inside bend of the first dipole magnet of the spectrometer. The collected charge was used in the calculation of absolute differential cross sections. However, because of inefficient current integration when running in dispersion-matched mode, a correction had to be applied. To determine this correction factor, data were also taken on a $^{26}$Mg target, with a thickness of 0.87 mg/cm$^{2}$. Cross sections for the $^{26}$Mg($^{3}$He,$t$) reaction were measured previously in an experiment ran in lower-resolution achromatic mode and used for the calibration of the unit cross sections \cite{ZEG06,ZEG07}. A 20\% correction to the cross sections for the present data had to be applied. Except for this normalization factor, angular distributions for the $^{26}$Mg($^{3}$He,$t$) reaction measured in the experiments employing dispersion-matched and achromatic tunes were consistent. The $^{26}$Mg($^{3}$He,$t$) spectrum was also useful for calibrating the triton energies measured in the spectrometer, since the excitation energy spectrum of $^{26}$Al is well known from two previous ($^{3}$He,$t$) experiments \cite{FUJ03,ZEG06} and other experimental studies \cite{END98}. A minor correction had to be applied due to the small difference in thickness for the $^{24}$Mg and $^{26}$Mg targets.
After this calibration, the uncertainties in the excitation energies of the $^{24}$Al spectrum were checked by using the $^{24}$Al ground state, the first excited state (at 0.4258(1) MeV; its location is well-known from a $\gamma$ decay measurement \cite{HON79}), two $^{16}$F states [$E_{x}$($^{16}$F)=0.193(6), 0.424(5) MeV] and the $^{12}$N ground state. The deviations between the extracted and known values were 5 keV or less. Taking into account the small uncertainties for correcting the energies due to recoil effects for reactions on different target nuclei, the uncertainties in the excitation energies of the $^{16}$F excited states and higher-order magnetic field abberations that affect the calibration, we assigned a minimum error of 10 keV to all excitation energies below 4 MeV in $^{24}$Al. Above 4 MeV, the energies become gradually more uncertain because there are few appropriate calibration levels in the $^{26}$Mg($^{3}$He,$t$) spectrum and no clear excitations from reactions on contaminants in the $^{24}$Mg target. Up to 5.5 MeV the assigned error was 20 keV and above 5.5 MeV 30 keV. The statistical and systematic errors in the peak fitting procedure were mostly less than 3 keV.

\begin{figure}
\includegraphics[scale=0.8]{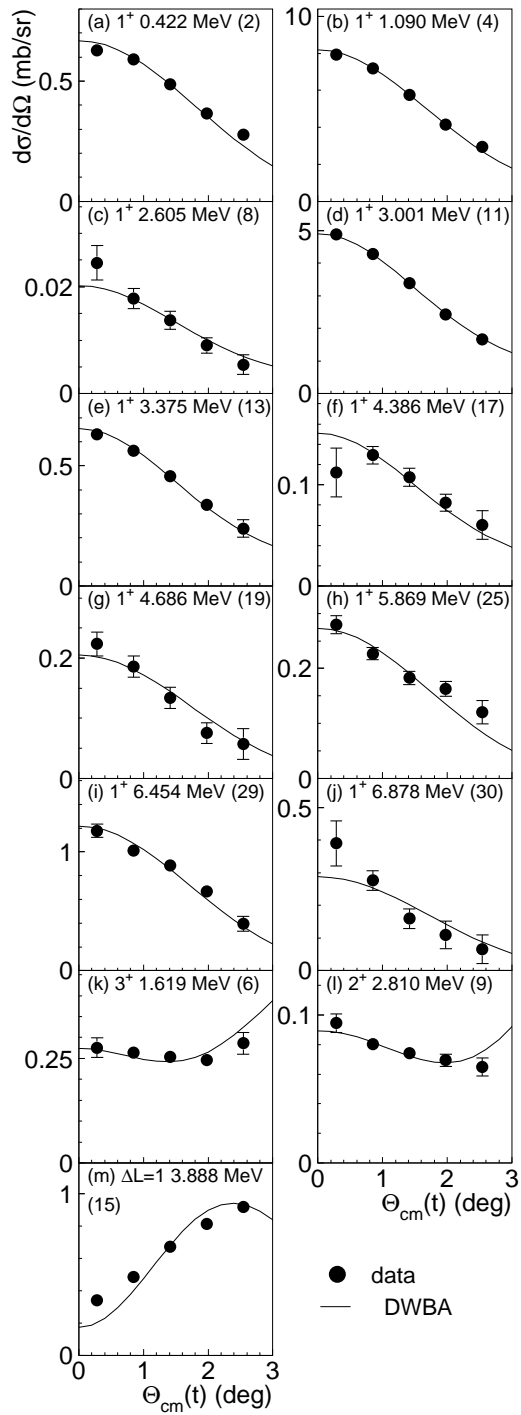}
\caption{\label{fig2}Differential cross sections for the $^{24}$Mg($^{3}$He,$t$) reaction at 420 MeV. (a-j) correspond to the GT transitions. (k) corresponds to the excitation of the $3^{+}$ state at 1.619 MeV, (l) corresponds to the excitation of the $2^{+}$ state at 2.810 MeV and (m) corresponds to the excitation of a likely dipole state at 3.888 MeV (its spin-parity is uncertain). All experimental differential cross sections are compared with DWBA calculations which were scaled to the data in a single-parameter fit (see text). The numbers in brackets correspond to the labels for the peaks in Fig. \ref{fig1}.}
\end{figure}

Fig. \ref{fig1} shows the energy spectrum of the $^{24}$Mg($^{3}$He,$t$) reaction up to $E_{x}$($^{24}$Al)=7 MeV. Most peaks are identified as excited states in $^{24}$Al and numbered. Certain excitations, for example peak 10, which in Fig. \ref{fig1} appears inseparable from peak 11, could only identified by inspecting the energy spectra at different scattering angles. Peaks due to the contaminants of significantly different mass in the target were easily identified from their kinematical shifts. Since the $^{12}$C($^{3}$He,$t$) and $^{16}$O($^{3}$He,$t$) spectra are well known experimentally, we could conclude that some of the broader states at higher excitation energies in $^{24}$Al were not a result of reactions on these contaminants. Some of the $^{24}$Al states are only weakly populated. These include states 7 and 8, which fall in the region of interest for the $^{23}$Mg($p,\gamma)^{24}$Al reaction. The kinematical shifts of these states matched  the expectation for a target with a mass close to A=24 and because their energies do not coincide with any known $^{16}$O($^{3}$He,$t$)$^{16}$F excitations or other possible contaminants. Likely contaminants would have produced strong signatures below the threshold of the $^{24}$Mg($^{3}$He,$t$) reaction ($Q$ value of -13.987 MeV). The most probable background reactions, $^{25}$Mg($^{3}$He,$t$) and $^{26}$Mg($^{3}$He,$t$), have Q values of -4.296 MeV and -4.023 MeV, respectively, and their spectra are well studied \cite{FUJ04c,FUJ03,ZEG06}. No significant signatures of the presence of these reactions were identified.

The data set was divided into five 0.5$^{\circ}$ (laboratory angle) angular bins. The yields for the peaks numbered in Fig. \ref{fig1} were obtained in each angular bin. If a peak was not isolated, the background under it was parameterized with a polynomial in the energy region close to the peak and a systematic error to the yield was assigned based on the ambiguity in estimating the background. If two or more peaks were not separated, fits were performed simultaneously for those peaks and a background included in the fit, if necessary.

GT states were identified by using their typical strongly forward-peaked differential cross sections, associated with angular momentum transfer $\Delta L=0$.\footnote{Since $^{24}$Mg has N=Z, no excitation of the isobaric analog state ($\Delta L=0$, $\Delta S=0$) is expected. A $T=2$ $0^{+}$ state has been found at 5.957 MeV \cite{BAN01} but it cannot be excited in the $\Delta T=1$ ($^{3}$He,$t$) reaction on $^{24}$Mg ($T=0$) } The differential cross sections of nearly all the numbered states in Fig. \ref{fig1} were compared with theoretical curves calculated in Distorted-Wave Born Approximation (DWBA) performed with the code \textsc{fold} \cite{FOLD}. The exceptions were peaks 5 and 10, because the systematic errors in the extraction of the differential cross sections were too large. In these cases, we could only confirm that they do not strongly peak at forward angles and hence are not associated with GT transitions.
The DWBA calculations were very similar to those discussed in Ref. \cite{ZEG06} where the $^{26}$Mg($^{3}$He,$t$) reaction at 420 MeV was studied. The structure input for the DWBA calculations, in the form of one-body transition densities, was calculated using the $sd$ shell-model interaction USDA \cite{BRO06} in proton-neutron formalism (isospin-nonconserving) with the code \textsc{OXBASH} \cite{OXBA}.

With the exception of peaks 5 and 10, all numbered peaks in Fig. \ref{fig1} were compared with the DWBA calculation for a state of the best-matching spin-parity and that was closest to the experimental excitation energy. For each state, the DWBA calculations were scaled by an angle-independent factor that was determined in a fit. In Figs. \ref{fig2}(a-j), the comparison between the scaled DWBA calculations and the data are shown for all states identified as GT transitions. To illustrate that the angular distribution for GT transitions can be uniquely identified, the differential cross sections and matched DWBA calculations are shown for the transitions to the $3^{+}$ state at 1.609 MeV [Fig. \ref{fig2}(k)] and the $2^{+}$ state at 2.790 MeV [Fig. \ref{fig2}(l)]. For both of these, the multipolarity is known \cite{FIR07}. In addition, the differential cross section for what is very likely a dipole transition at 3.862 MeV is shown in Fig. \ref{fig2}(m). The DWBA calculation shown in the plot assumes a transition to a $2^{-}$ state. Given the limited angular range of the data set, it was not possible to unambiguously assign the multipolarity for positive parity transitions with $\Delta L > 0$ or for negative parity states of various total angular momentum transfer. In the discussion below, the only distinction made is, therefore, between $\Delta L=0$ (GT) transitions and transitions with $\Delta L\neq0$. It was not possible to unambiguously identify GT strengths at excitation energies above 7 MeV in the spectrum, perhaps partially due to the limited angular coverage. Small amounts of GT strengths above 7 MeV were identified in the $^{24}$Mg($p,n$) data \cite{AND91} (see also below).

For the states identified as having $J^{\pi}=1^{+}$, the zero-degree cross section was extracted from the fitted theoretical curve, with the uncertainty deduced from the fitting error. To extract the GT strength [$B$(GT)] for each state in the eikonal approximation \cite{TAD87,ZEG07} knowledge of the unit cross section ($\hat{\sigma}$) and the differential cross section at zero momentum transfer ($q=0$) is required:
\begin{equation}
\label{eq:eik}
B(GT)=\frac{d\sigma}{d\Omega}(q=0)/\hat{\sigma},
\end{equation}
The differential cross sections at $q=0$ were obtained by extrapolating the data at finite $q$ (i.e. at finite $Q$ value and $0^{\circ}$ scattering angle) to $q=0$ (i.e. $Q=0$ and $0^{\circ}$) by using the DWBA results:
\begin{equation}
\label{eq:extra}
\frac{d\sigma}{d\Omega}(q=0)=\left[\frac{\frac{d\sigma}{d\Omega}(q=0)}{\frac{d\sigma}{d\Omega}(Q,0^{\circ})}\right]_{\text{DWBA}}\times\left[\frac{d\sigma}{d\Omega}(Q,0^{\circ})\right]_{\text{exp}}.
\end{equation}
In this equation, `$\text{DWBA}$' refers to calculated values in the DWBA code. The unit cross section $\hat{\sigma}$ was calculated with \cite{ZEG07}:
\begin{equation}
\label{eq:unit}
\hat{\sigma}=109\times A^{-0.65}.
\end{equation}
For the cases studied in Ref. \cite{ZEG07}, the error when using this equation was less than about 5\%, except for certain transitions in which interference effects between $\Delta L=0$ and $\Delta L=2$  amplitudes to the excitation of $J^{\pi}=1^{+}$ GT states were strong. The $\Delta L=2$ contributions are mediated via the tensor-$\tau$ component of the effective nucleon-nucleon interaction \cite{LOV81,LOV85}. The uncertainties associated with this interference were studied in detail in Refs. \cite{COL06,FUJ07,ZEG06,ZEG07,ZEG08}. In a theoretical study of the $^{26}$Mg($^{3}$He,$t$) reaction \cite{ZEG06} based on DWBA calculations using $sd$ shell-model one-body transition densities (with the USDB interaction) and the Love-Franey effective nucleon-nucleon interaction \cite{LOV81,LOV85}, it was shown that errors associated with the interference between $\Delta L=0$ and $\Delta L=2$ amplitudes increase with decreasing values of $B$(GT). For $B$(GT)=1, 0.1, 0.01 and 0.001 the estimated uncertainties were 3\%, 11\%, 20\% and 28\%, respectively (see Fig. 6 and Eq. (7) in Ref. \cite{ZEG06}). An identical procedure for estimating the errors due to the tensor-$\tau$ interaction in the case of the $^{24}$Mg($^{3}$He,$t$) reaction was performed for the present work and very similar results were found.
Since in the following, we will compare data from ($^{3}$He,$t$) and ($p,n$) reactions, it should be noted that the tensor-$\tau$ interaction also introduces uncertainties for the latter \cite{TAD87}. However, as shown in Ref. \cite{COL06} for the case of a $^{58}$Ni target, for a given transition, the effects of the interference are not necessarily similar in magnitude for the ($p,n$) and ($^{3}$He,$t$) reactions; even the sign of the interference can be different. Since it is hard to estimate the magnitude on a level-by-level basis, the uncertainties due to the tensor-$\tau$ interaction will not be quoted explicitly in the following section, but must be kept in mind when comparing data sets and checking the validity of theoretical calculations.

\begin{table*}
\caption{\label{tab:table1}Experimental results from the $^{24}$Mg($^{3}$He,$t$) experiment at 420 MeV (columns 1-5) and the comparison with $^{24}$Mg($p,n$) results \cite{AND91} for the extraction of GT strength (colums 6-7). In addition, results for the excitation energies from three $^{24}$Mg($^{3}$He,$t$) experiments at beam energies of 81 MeV \cite{GRE91}, 60 MeV \cite{KUB95} and 30 MeV \cite{VIS07} are shown (columns 8-10) up to the region of interest for the $^{23}$Mg(p,$\gamma$) reaction at astrophysical temperatures.}
\begin{ruledtabular}
\begin{tabular}{llllllllll}
 \multicolumn{5}{c}{present data} & \multicolumn{2}{c}{(p,n) \cite{AND91}} &($^{3}$He,$t$) \cite{GRE91} &  ($^{3}$He,$t$) \cite{KUB95} & ($^{3}$He,$t$) \cite{VIS07} \\
\cline{1-5} \cline{6-7} \cline{8-10}
Fig. 1 & $E_{x}$($^{24}$Al) & $\Delta L\footnotemark[1]$ & $d\sigma/d\Omega(0^{\circ})$\footnotemark[2] & $B$(GT)\footnotemark[2]\footnotemark[3] & $E_{x}$($^{24}$Al) & $B$(GT)\footnotemark[4] & $E_{x}$($^{24}$Al) & $E_{x}$($^{24}$Al) & $E_{x}$($^{24}$Al) \\
label & (MeV) &  & (mb/sr) &  & (MeV) &  & (MeV) & (MeV) & (MeV) \\
\hline
1 & 0 & $\neq 0$ & - & - &  &  & 0 & 0 &   \\
2 & 0.422(10)\footnotemark[5] & 0 & 0.67(1) & 0.054(1) & 0.44\footnotemark[5] & 0.050(1) & 0.439(6)\footnotemark[5] & 0.432(10)\footnotemark[5] &  \\
3 & 0.492(10) & $\neq 0$ & - & - &  &  & 0.511(4) & 0.506(10) &  \\
4 & 1.090(10) & 0 & 8.18(3) & 0.668(3) & 1.07 & 0.613(2) & 1.111(3) & 1.101(10) &  \\
 &  &  &  &  &  &  & 1.275(5) & 1.260(10) &  \\
5 & 1.555(10) & $\neq 0$ & - & \footnotemark[6] & 1.58 & 0.020(6) & 1.563(7) & 1.535(10) & 1.543(6) \\
6 & 1.619(10) & $\neq 0$ & - & - &  &  & 1.638(8) & 1.614(10) & 1.619(6) \\
7 & 2.349(10) & $\neq 0$ & - & - &  &  & 2.369(4) & 2.328(10) & 2.346(6) \\
 &  &  &  &  &  &  & 2.546(7) & 2.521(10) & 2.524(6) \\
8 & 2.605(10) & 0 & 0.020(2) & 0.0017(2) &  &  &  &  &  \\
9 & 2.810(10) & $\neq 0$ & - & - &  &  & 2.823(6) & 2.787(10) & 2.792(6) \\
10 & 2.89(20) & $\neq 0$ & - & - &  &  & 2.920(23) & 2.876(10) & 2.874(6) \\
11 & 3.001(10)\footnotemark[7] & 0 & 4.90(3) & 0.416(3) & 2.98 & 0.362(5) & 3.037(16) & 3.002(10) & 2.978(6) \\
 &  &  &  &  &  &  &  &  & 3.019(6) \\
12 & 3.292(10) & $\neq 0$ & - & - &  &  & ... & ... & ... \\
13 & 3.375(10)\footnotemark[7] & 0 & 0.65(1) & 0.056(1) & 3.33 & 0.059(1) &  \multicolumn{3}{c}{additional states}  \\
14 & 3.691(10) & $\neq 0$ & - & - &  &  & \multicolumn{3}{c}{not included in table} \\
15 & 3.888(10) & $\neq 0$ & - & - &  &  &  &  &  \\
16 & 4.088(50) & $\neq 0$ & - & - &  &  &  &  &  \\
17 & 4.386(20)\footnotemark[7] & 0 & 0.15(1) & 0.013(1) &  &  &  &  &  \\
18 & 4.426(20) & $\neq 0$ & - & - &  &  &  &  &  \\
19 & 4.686(20)\footnotemark[7] & 0 & 0.20(3) & 0.018(3) & 4.69 & 0.015(4) &  &  &  \\
20 & 4.734(20) & $\neq 0$ & - & - &  &  &  &  &  \\
21 & 4.971(20) & $\neq 0$ & - & - &  &  &  &  &  \\
22 & 5.312(20) & $\neq 0$ & - & - &  &  &  &  &  \\
23 & 5.483(20) & $\neq 0$ & - & - &  &  &  &  &  \\
24 & 5.692(30) & $\neq 0$ & - & - &  &  &  &  &  \\
25 & 5.869(30)\footnotemark[7] & 0 & 0.27(2) & 0.024(2) &  &  &  &  &  \\
26 & 5.952(30) & $\neq 0$ & - & - &  &  &  &  &  \\
27 & 6.141(30) & $\neq 0$ & - & - &  &  &  &  &  \\
28 & 6.214(30) & $\neq 0$ & - & - &  &  &  &  &  \\
29 & 6.454(30) & 0 & 1.22(3) & 0.112(3) & 6.46 & 0.068(1) &  &  &  \\
30 & 6.878(30) & 0 & 0.3(1) & 0.03(1) &  6.87 & 0.029(1) &  &  &  \\
31 & 6.896(30) & $\neq 0$ & - & - &  &  &  &  &  \\ \hline
 &  &  & $\Sigma B$(GT) & 1.39(1) &  & 1.216(9) &  &  &  \\
 &  &  &  &  & $>7$ & 0.084(9) &  &  &  \\
 &  &  &  $\Sigma B$(GT)&  &  & 1.300(13) &  &  &  \\
\end{tabular}
\end{ruledtabular}
\footnotetext[1]{All states which were not clearly related to $\Delta L=0$ transitions were assigned $\Delta L\neq 0$, even though in most cases a reasonable judgment on the angular momentum transfer can be made (see discussion in text).}
\footnotetext[2]{Listed errors are due to statistical and fitting uncertainties only. Systematic errors are 10\% (see text).}
\footnotetext[3]{$B$(GT)=$\frac{d\sigma}{d\Omega}(q=0)/(109\times(24)^{-0.65})$ (see text).}
\footnotetext[4]{Uncertainties are calculated from the error bars given in \cite{AND91} for the differential cross sections at $0^{\circ}$ and represent statistical and fitting uncertainties only.}
\footnotetext[5]{Corresponds to the 0.4258(1) state for which the energy is well known from $\gamma$ decay \cite{HON79}.}
\footnotetext[6]{Upper limit for $B$(GT)=0.005, assuming all events in this peak are due to a GT transition, ignoring possible contamination from the excitation of the $^{16}$F(g.s.) and the non-matching angular distribution (see text). }
\footnotetext[7]{Likely corresponds to a $1^{+}$ state observed in the $\beta$-delayed proton decay of $^{24}$Si \cite{BAN01} (see text).}
\end{table*}

\section{Results and discussion.}
\label{sec:results}
\subsection{GT strengths}
\label{sec:gt}
In Table \ref{tab:table1}, the results of the experiment are summarized and compared with previous results for the GT strength distribution extracted from the $^{24}$Mg($p,n$) reaction at 135 MeV \cite{AND91} with a resolution of 310 keV. The listed uncertainties for the extracted GT strengths for both data sets include statistical and fitting errors only. For the present data, the combined error related to the uncertainty in the unit cross section (5\%) and the uncertainty in the $^{24}$Mg target thickness was estimated to be 10\%. The same value was given for the error in the extraction of the $B$(GT) from the ($p,n$) data \cite{AND91}.

Overall a good correspondence between the GT strength from the present data and the $^{24}$Mg($p,n$) data is found, but there are a few discrepancies. In Ref. \cite{AND91} a $1^{+}$ state was reported at 1.58 MeV, with a $B$(GT) of 0.02, presumably corresponding to the state at 1.555 MeV (peak 5) in the present data. As mentioned above, we found that the angular distribution of this state is not associated with $\Delta L=0$, although the analysis is complicated by the contamination from the $^{16}$O($^{3}$He,$t$)$^{16}$F(g.s.) reaction. However, even if this contamination is ignored and it is assumed that all events in this peak are due to a $1^{+}$ state in $^{24}$Al, the $B$(GT) would be 0.005. This upper limit is far below the value reported in Ref. \cite{AND91}. It was, therefore, concluded that this is not a $1^{+}$ state.

\begin{figure}
\includegraphics[width=7.5cm]{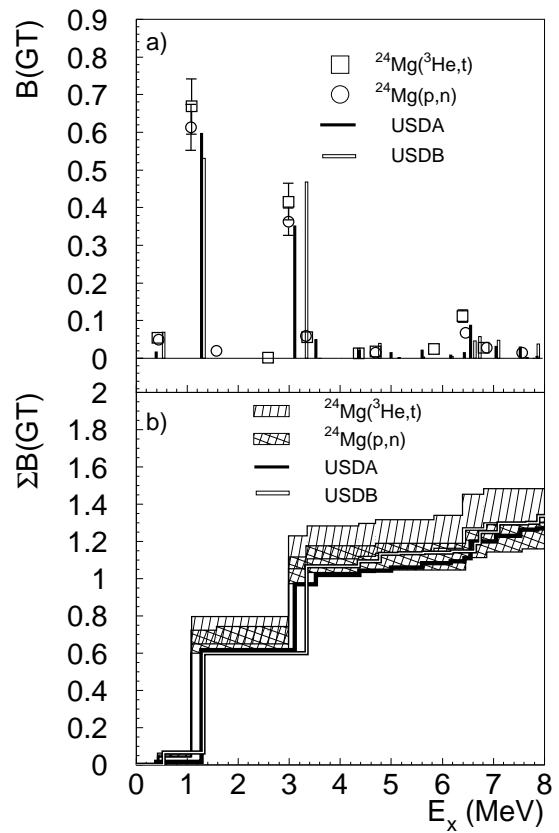}
\caption{\label{fig3}(a) Comparison of GT strength distributions extracted from the present $^{24}$Mg($^{3}$He,$t$) data, the $^{24}$Mg($p,n$) data \cite{AND91} and shell-models calculations using the USDA and USDB interactions \cite{BRO06}. (b) Cumulative sums of strengths for both data sets and theoretical calculations. The widths of the bands for the experimental results represent the combination of statistical and systematic uncertainties.}
\end{figure}

The $1^{+}$ state found in the present data at 2.605 MeV (peak 8 in Fig. \ref{fig1}) was not seen in the $(p,n)$ data, but that is understandable from the very small strength [0.0017(2)] associated with this excitation and from the fact that the resolution in the $(p,n)$ experiment was 310 keV, compared to the 35 keV reported here. Because the cross section for the excitation of this state is so small, one should perhaps worry about significant contributions from multistep processes even at the beam energy of 420 MeV. Such non-direct contributions can affect the angular distribution \cite{IGA78}, and thus the identification of the state.
Two other $1^{+}$ states found at 4.386 MeV and 5.869 MeV were not seen in the $(p,n)$ data. These states likely correspond to $1^{+}$ states measured at 4.40 MeV and 5.76 MeV in the measurement of $\beta$-delayed proton decay of $^{24}$Si $\beta$ \cite{BAN01} (Ref. \cite{CZA98} reports the former of these two at 4.38(5) MeV). Several of the other $1^{+}$ states found in both the $(p,n)$ and ($^{3}$He,$t$) experiments correlate to $1^{+}$ states found in those data, as indicated in Table \ref{tab:table1}.

In Fig. \ref{fig3}(a), the extracted GT strengths from the $(p,n)$ and present ($^{3}$He,$t$) experiment are plotted, together with the results of the shell-model calculations. The energy axis has been cut off at 8 MeV, but as shown in Table \ref{tab:table1}, in the analysis of the ($p$,$n$) data small amounts of GT strength were detected up to about 11 MeV. For both data sets, the above-mentioned uncertainties of 10\% were used in combination with the statistical errors to calculate the total error bars.
Besides the strength distribution calculated with the above-mentioned USDA interaction, the results using the USDB interaction \cite{BRO06} in the $sd$ shell-model space are also shown. The difference between the USDA and USDB lies in the number of varied linear combinations of Hamiltonian parameters in the construction for each interaction; The USDA Hamiltonian was more constrained than the USDB Hamiltonian. Both theoretical calculations have been multiplied by a factor of 0.59 \cite{WIL83,BRO85,BRO88} (the error in this factor is 0.03), to take into account the quenching of the strength due to a combination of configuration mixing with $2p-2h$ states \cite{HYU80,ARI99,YAK05} and coupling to the $\Delta(1232)$-isobar nucleon-hole state \cite{ERI73}.
Given the uncertainties (including those due to the tensor-$\tau$ interaction, although they are not included in the plot), the two data sets and the theoretical calculations agree well. The calculation with the USDA interaction does slightly better in predicting the strength distribution than the USDB interaction. For example, with the USDA interaction the splitting of the GT strength into a strong and weaker state near 3.5 MeV matches well with the experimental results, whereas the calculation using the USDB interaction predicts a single $1^{+}$ state in this region. At higher excitation energies, the USDA interaction also does slightly better in predicting the location of individual states. Neither calculations predict a $1^{+}$ state near 2.5 MeV with small B(GT) as observed in the present data.

In Fig. \ref{fig3}(b), the cumulative sums of GT strengths for the data sets and the theoretical calculations are shown. Slightly more strength is found in the present data, compared to the $(p,n)$ results (see also Table \ref{tab:table1}). However, given the uncertainties, on the whole the data sets match well and both the theoretical calculations reproduce the experimental strength distributions.

\subsection{Excitation energies of low-lying states in $^{24}$Al}
\label{sec:levels}
In the last three columns of Table \ref{tab:table1}, the excitation energies found in the three $^{24}$Mg($^{3}$He,$t$) experiments performed at beam energies of 81 MeV \cite{GRE91}, 60 MeV \cite{KUB95} and 30 MeV \cite{VIS07} are given, up to $E_{x}=3$ MeV, which includes range above the proton separation energy that is important for the astrophysical applications. Above 3 MeV, a state-by-state comparison between the low-energy data and the results from the present experiment is difficult because of the increasing level density and the difference in sensitivities for transitions of various angular momentum transfers. Note that in the most recent low-energy experiment \cite{VIS07} energy levels below 1.5 MeV were not measured. As discussed in the introduction, there are significant differences between the three previous ($^{3}$He,$t$) measurements: the energy levels from Ref. \cite{GRE91} are inconsistent with those of Refs. \cite{KUB95,VIS07}.

\begin{figure}
\includegraphics[width=7.5cm]{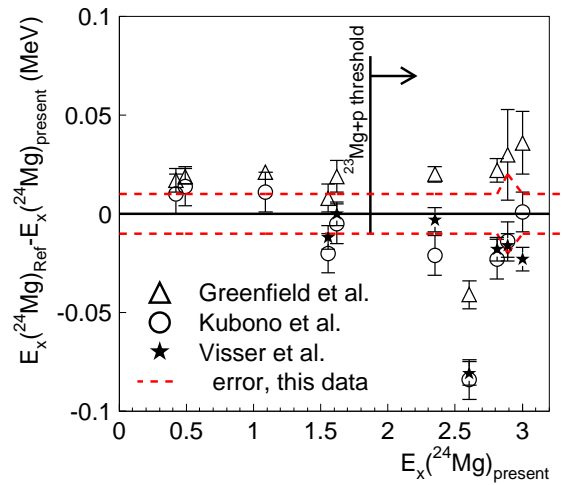}
\caption{\label{fig4}Comparison between excitation energies for low-lying states in $^{24}$Al for the ($^{3}$He,$t$) experiments by Greenfield et al. \cite{GRE91}, Kubono et al. \cite{KUB95}, Visser et al. \cite{VIS07} and the present data. The present data are taken as the reference and correspond to a value of 0 on the vertical axis. The dashed lines (color online) correspond to the error margins in the present data. The present data are also used for the horizontal-axis scale, assuming the presence of a matched state in the other data. For the state found at 2.605 MeV in the present data, a significant discrepancy with the assumed matching state in the other experiments is found, pointing to the presence of a previously unknown state (see text).}
\end{figure}

In Fig. \ref{fig4}, the energies of low-lying levels measured in the previous ($^{3}$He,$t$) experiments and the present data are compared. The energies from the present experiment are chosen as reference. The following observations are made. The levels extracted in Ref. \cite{GRE91} lie systematically higher than those the present data, by 15-20 keV, except for one state at 2.605 MeV. Ignoring that state for reasons discussed below, a $\chi^{2}$ test for the consistency between the data from Ref. \cite{GRE91} and the present data showed that it is outside the 99\% confidence interval ($\chi^{2}=23.5$ with 9 degrees of freedom). Consistency tests between the present data and the data from Ref. \cite{KUB95} ($\chi^{2}=9.46$ with 9 degrees of freedom) and Ref. \cite{VIS07} ($\chi^{2}=7.98$ with 6 degrees of freedom) were well within the 95\% confidence interval. If the results from Ref. \cite{GRE91} are rejected, the weighted average of the results for the first and most important excited state above the proton-capture threshold from Refs. \cite{KUB95,VIS07} and the present data is 2.343(5) MeV. This is consistent with the value used in Ref. \cite{VIS07} to calculate the proton-capture rate.

The state found at 2.605 MeV is about 80 keV higher than its assumed counterparts ($\sim 2.524$ MeV) in the works of Refs. \cite{KUB95,VIS07}. The deviation is less (40 keV) when comparing with the results from Ref. \cite{GRE91}, but after taking into account that the energies in that data are systematically higher, it is almost equally inconsistent. We identified the state at 2.605 MeV as a $1^{+}$ state (although somewhat uncertain, due to its small cross section and possible contributions from non-direct production mechanisms) based on its angular distribution. Transitions with a large angular momentum transfer are relatively strongly excited in the low-energy experiment and strongly suppressed at 420 MeV (at forward scattering angles). On the other hand $\Delta L=0$ transitions are easier to identify at the higher energy and the 2.605-MeV state could thus have been missed in the experiments at the lower beam energies.
We, therefore, conclude that the state at 2.605 MeV seen in the present data is not the same as the state at $\sim 2.524$ MeV seen in the low-energy ($^{3}$He,$t$) data. Hence, it should in principle be included separately into the calculation of the $^{23}$Mg($p$,$\gamma$) reaction rate as the third state above the capture threshold.

Since no $1^{+}$ state is predicted near this excitation energy in the shell-model calculations, one cannot calculate the spectroscopic factor $S$ (needed for calculating the proton partial width $\Gamma_{p}$) and reduced $\gamma$-ray transition strengths (needed for calculating the $\gamma$-ray width $\Gamma_{\gamma}$), as was done in Refs. \cite{HER98,VIS07} for other the other transitions just above the proton separation energy.
It is noted that in the shell-model calculations presented in Table I of Ref. \cite{HER98} a $3^{+}$ state is predicted at 2.629 keV, but connected to an state experimentally observed at 2.900 MeV (Ref. \cite{VIS07} uses 2.874 MeV for this state). The calculations in Ref. \cite{HER98} were performed with the original USD interaction \cite{WIL84}. The locations of the $J^{\pi}=1-5^{+}$ levels below 3 MeV as calculated with the USDA and USDB interactions \cite{BRO06} used in this work are consistent with the USD values to within 190 keV. The USDA(USDB) interaction situate this particular $3^{+}$ state at 2.591 MeV (2.605 MeV). One can speculate that the new state at 2.605 MeV is in fact the $3^{+}$ state predicted at that energy in the shell-model calculations, but we stress that this is unlikely based on the measured angular distribution and would require strong multiple-step contributions\footnote{ In the one-step DWBA, the calculated cross section for $3^{+}$ shell-model state at 2.605 MeV is a factor of 30 smaller than the cross section for the $3^{+}$ state at 1.619 MeV. By using the measured cross section of latter state, one thus expects a cross section of about 0.008 mb/sr at $0^{\circ}$ for the state at 2.605 MeV, a factor of 3 below the experimental value.}. Nevertheless, based on this speculation, and following Ref. \cite{HER98}, we calculated $\Gamma_{\gamma}$ and $\Gamma_{p}$ in the shell model and deduced the resonance strength $\omega\gamma=10$ meV using the USDB interaction\footnote{Ref \cite{HER98} used the original USD interaction \cite{WIL84} and found 12 meV}. At a resonance energy of 0.734 MeV, the inclusion of this state would only change the proton-capture rate on $^{23}$Mg by maximally 2.5\% at 2$\times10^{9}$ K, which is much smaller than other uncertainties in the rate given in Ref. \cite{VIS07}. The reason is that its resonance strength is much smaller than the strengths of the first two states above threshold. To do a more detailed and reliable calculation and to ensure that the impact of the new state at 2.605 MeV on the astrophysical rate calculation is indeed small, confirmation of its nature is desirable.

\section{Conclusion}
We have measured the $^{24}$Mg($^{3}$He,$t$) reaction at 420 MeV and used the empirical relationship for the unit cross section as a function of mass number for this probe to extract the Gamow-Teller strengths for transitions to $1^{+}$ states in $^{24}$Al. Owing to the high energy resolution achieved, several new small GT states have been discovered at energies below 7 MeV. Nevertheless, taking into account the uncertainties involved with the extraction of GT strengths using fitted trends of unit cross section, the present ($^{3}$He,$t$) and previous $(p,n)$ experiments are in good agreement. The experimental results were also compared with theoretical calculations employing the USDA and USDB interactions and a satisfactory consistency was observed.

Because of the high energy resolution achieved, the excitation energies of several levels of importance for estimating the proton-capture rate on $^{23}$Mg for astrophysical purposes could be studied. The results were consistent with two previous ($^{3}$He,$t$) experiments performed at beam energies of 60 MeV and 30 MeV, giving further indication that the values extracted from a third ($^{3}$He,$t$) experiment (at 81 MeV) are systematically too high. However, a new state is identified at 2.605(10) MeV, which is 0.734 MeV, and thus the third state, above the proton capture threshold. Based on the comparison with DWBA calculations, and from the fact that it was not observed in the low-energy experiments, we tentatively identified this as a $1^{+}$ state. This assignment is somewhat uncertain because of the small cross section and the associated possibility that non-direct contributions might have affected the angular distribution. Speculating it to be a $3^{+}$ state, as predicted in shell-models, the proton-capture rates calculated in previous works are not strongly affected. However, more experimental information on this new state is needed to better judge the impact on the $^{23}$Mg($p,\gamma$) reaction in stellar environments.

\begin{acknowledgments}
We thank the cyclotron staff at RCNP for their support during the experiment. This work was supported by the US NSF (PHY0216783 (JINA), PHY-0555366, PHY-0606007), the
Ministry of Education, Science, Sports and Culture of Japan (Grant No. 18540270) and the DFG, under contract No. Br 799/12-1.
\end{acknowledgments}

\bibliography{prc}

\end{document}